\documentclass[final,authoryear,12pt,5p,times]{elsarticle}

\usepackage{bm}
\newcommand{\bmath}[1]{\bm{#1}}

\newcommand{\mathbmss}[1]{\bm{\mathsf{#1}}}
 \usepackage{graphicx}

\usepackage{amssymb}
\usepackage{amsmath}

\DeclareMathOperator\arsinh{arsinh}
\DeclareMathOperator\arcosh{arcosh}

\journal{Astronomy and Computing}

\begin{document}
\sloppy

\begin{frontmatter}


%

\title{Fast error--safe MOID computation involving hyperbolic orbits}


\author{Roman V. Baluev}
\address{Saint Petersburg State University, Faculty of Mathematics and Mechanics, Universitetskij
pr. 28, Petrodvorets, Saint Petersburg 198504, Russia}
\address{Central Astronomical Observatory at Pulkovo of the Russian Academy of Sciences,
Pulkovskoje sh. 65/1, Saint Petersburg 196140, Russia}
 \ead{r.baluev@spbu.ru}

\begin{abstract}
We extend our previous algorithm computing the minimum orbital intersection distance (MOID)
to include hyperbolic orbits, and mixed combinations ellipse--hyperbola. The MOID is
computed by finding all stationary points of the distance function, equivalent to finding
all the roots of an algebraic polynomial equation of $16$th degree. The updated algorithm
carries about numerical errors as well, and benchmarks confirmed its numeric reliability
together with high computing performance.
\end{abstract}


\end{frontmatter}



\section{Introduction}
The MOID parameter, or the minimum distance between points on two Keplerian orbits, is an
important practical tool measuring the closeness of two Keplerian trajectories in the
$\mathbb R^3$ space. This parameter is frequently used in studies of Potentially Hazardous
Objects (PHOs) and Near-Earth Objects (NEOs), see e.g. \citep{Sitarski68,Dybczynski86}.
Also, the MOID and MOID-like quantities can be used to estimate possible visibility
condition of one object from another \citep{GronchiNiederman20}.

The MOID is a minimum of some distance or distance-like function $\rho(u,u')$ that depends
on two arguments, determining positions on two orbits. Multiple methods of finding the
minima of $\rho(u,u')$ are available
\citet{KholshVas99,Gronchi02,Baluev05,Gronchi05,Armellin10,Hedo18,BalMik19}, as well as
methods allowing to put useful bounds on the MOID or related quantities
\citep{MikBal19,GronchiNiederman20}.

The fastest methods appear those in which both $u$ and $u'$ are solved for rather than
found by numeric optimization \citep{KholshVas99,Gronchi02,Gronchi05,Baluev05,BalMik19}.
The task is analytically reduced to solving a nonlinear equation with respect to $u$ and
then expressing $u'$ also analytically. However, methods of this class are relatively
vulnerable with respect to numeric uncertainties through loosing real roots in nearly
degenerate cases. This effect appears because the equation for $u$ is quite complicated and
has algebraic degree of $16$ in the general case. It often have close (almost multiple)
roots that are always difficult for numeric processing.

The recent work by \citet{BalMik19} represents an efficient solution of this issue based on
careful treatment of numeric errors appearing on the way. Basically, it represent a numeric
implementation of the algebraic approach presented by \citet{KholshVas99}, similar to the
one presented by \citet{Gronchi02,Gronchi05}. However, this newer algorithm is only capable
to process elliptic orbits. This appears because the arguments $u$ and $u'$ have the
meaning of eccentric anomaly that is not sensible on a hyperbolic orbit. This contrasts
with \citep{Gronchi05} where this issue does not appear because they used true anomalies as
free variables. However, that code revealed numeric unreliability on some orbit pairs
\citep{Hedo18,BalMik19}. Though the fraction of such cases is small, they often escape from
being self-diagnosed, making such code less prefered than alternatives that avoid failures.
Simultaneously, the Gronchi code appears somewhat slower than the alternatives from that
works.

In this work we have a goal to extend the numerically stable and fast code from
\citep{BalMik19}, making it capable to process hyperbolic orbits as well. This relies on
the work by \citet{Baluev05}, where the analytic theory from \citep{KholshVas99} was
extended to all types of Keplerian orbits in any combinations. The need of processing the
hyperbolic orbits is highlighted by recent discoveries of new interstellar objects passing
close to Sun. In particular, the cometary object 2I/Borisov has an orbital eccentricity
above $3$, and its trajectory passes through the inner part of Solar system
\citep{Guzik19}. Moreover, early orbital solutions proposed that its could be a NEO. It
was thus listed on IAU's Minor Planet Center's NEO Confirmation Page as gb00234. Therefore,
one may consider the task of quick computation of all MOIDs between such an object and Main
belt orbits with a goal to reveal asteroids that have a risk of collision or close
approach. Small MOID does not guarantee such an event in itself, as the objects would need
to also appear with proper orbital positions for a collision. But this approach might help
to filter away those asteroids that do not have that chance at all.

The C++ source code of our MOID library named {\sc distlink} is available for download at
\texttt{http://sourceforge.net/projects/distlink}.

The structure of the paper is as follows. In Sect.~\ref{sec_math}, we discuss the
mathematical framework for MOIDs of hyperbolic orbits. Sect.~\ref{sec_alg} describes the
computing algorithm. Sect.~\ref{sec_perf} presents its performance tests.

\section{Mathematical setting}
\label{sec_math}
Consider two confocal orbits: $\mathcal O$ determined by the five geometric Keplerian
elements $a,e,i,\Omega,\omega$, and $\mathcal O'$ determined analogously by the same
variables with a stroke. We need to compute the minimum of the distance $|\bmath r - \bmath
r'|$ between two points lying on the corresponding orbits, and the orbital positions where
this minimum is attained. In this work we consider elliptic $\mathcal E$ as well as
hyperbolic $\mathcal H$ orbits, thus we have four possible combinations: $\mathcal
E\mathcal E$, $\mathcal E\mathcal H$, $\mathcal H\mathcal E$, $\mathcal H\mathcal H$. Here
we do not consider parabolic cases, as they appear degenerate both from the elliptic or
hyperbolic point of view, resulting in reduced degree of the main polynomial. They require
a special treatment therefore.

In the purely elliptic case the method by \citet{KholshVas99} reduced the problem to
solving for the roots of a trigonometric polynomial $g(u)$ of minimum possible algebraic
degree $16$ (trigonometric degree $8$). This polynomial can be expressed through the
Keplerian elements of $\mathcal O$ and $\mathcal O'$, and the associated formulae are
omitted here for brevity.

For each root $g(u)$ we can determine the second position $u'$ from explicit equations. In
non-degenerate cases there is only a single value of $u'$ that corresponds to a particular
solution for $u$.

Finally, after both the orbital positions $u$ and $u'$ were determined, the squared
distance between these points is scaled as $\rho(u,u') = |\bmath r - \bmath r'|^2 /
(2aa')$. The adimentional function $\rho$ can be used to compare different solutions (roots
of $g(u)$).

This central function $g$ can be rewritten in the standard trigonometric form:
\begin{equation}
g(u) = \sum_{k=-N}^N c_k {\mathrm e}^{iku},
\label{gcan}
\end{equation}
where $N=8$. By making the substitution $z={\mathrm e}^{iu}$ or $w={\mathrm e}^{-iu}$, we
can transform it to:
\begin{equation}
g(u) = \sum_{k=-N}^N c_k z^k = \mathcal P(z) w^N = \mathcal Q(w) z^N.
\label{gPQ}
\end{equation}
The task of finding roots of $g(u)$ becomes equivalent to solving the algebraic equation,
$\mathcal P(z)=0$ or $\mathcal Q(w)=0$. All complex roots of $g(u)$ combine into conjugate
pairs in terms of $u$, corresponding to the relationship $z\mapsto 1/z^*$ in terms of $z$.
Only $z$ with unit absolute value correspond to real $u$.

All coefficients $c_k$ can be formally expressed through Keplerian elements, but in
practice this does not appear possible even using computer algebra. Still, a short explicit
form for $c_{\pm 8}$ is given in \citep{BalMik19}.

Cases when one of the orbits is hyperbolic or parabolic were considered in
\citep{Baluev05}. In the hyperbolic case, we should make a replacement
\begin{equation}
u \mapsto i u, \quad \sqrt{1-e^2} \mapsto i \sqrt{e^2-1},
\end{equation}
where $i$ is the imaginary unit (not to be mixed with the inclination), and ``new'' $u$
attains the meaning of the hyperbolic analogue of eccentric anomaly.

The entire set of resulting formulae is given in \citep{Baluev05}, and we do not replicate
them all here. We only notice the following. In the $\mathcal E\mathcal H$ case the
polynomial $g$ preserves its usual trigonometric form~(\ref{gcan}), so the computing
algorithm remains nearly unchanged. However, if the first orbit is hyperbolic, $g$ becomes
\begin{equation}
g(u) = \sum_{k=-N}^N c_k {\mathrm e}^{-ku},
\label{gcanh}
\end{equation}
and it is reduced to the algebraic form~(\ref{gPQ}) with $z={\mathrm e}^{-u}$ or
$w={\mathrm e}^u$. Contrary to the elliptic case, all coefficients $c_k$ are real now, and
$z$ should be real. Complex roots combine into conjugate pairs (in terms of $u$ as well as
$z$).

In the elliptic case the number of real roots of $g(u)$ cannot be smaller than $4$, taking
the multiplicity into account, because $\rho(u,u')$ is a continuous function defined on a
torus \citep{KholshVas99}. This appears because we should necessarily have at least one
minimum and at least one maximum of $\rho$, and then the number of saddle points in a
general (non-degenerate) case should be at least $2$ following the Morse theory
\citep{Gronchi05}. The latter says that No. of maxima + No. of minima - No. of saddles =
Euler--Poincar\'e characteristic (EPC) of the manifold (${\rm EPC}=0$ for torus). However,
when one or both the orbits are hyperbolic, function $\rho$ is defined on a non-compact
domain, either cylinder (${\rm EPC}=0$) or plane (${\rm EPC}=1$). Also, we should treat
each hyperbola branch separately, so the total number of critical points is doubled or
quadrupled. In a mixed ($\mathcal E\mathcal H$ and $\mathcal H\mathcal E$) case for each
hyperbola branch we have at least one minimum, and hence at least one saddle point from the
Morse theory, or $2$ critical points per branch and $4$ ones in total. In the $\mathcal
H\mathcal H$ case we have at least $4$ local minima (the number of possible branch
combinations), and no informative limit on saddle points. Therefore, in all non-degenerate
cases (no root multiplicity) we have at least $4$ critical points in any orbital
combination. Simultaneously, this number should be even, because complex roots always
combine into pairs and the algebraic degree of $g$ is always even.

If at least $\mathcal O$ is hyperbolic then there should be even number of roots \emph{per
each hyperbola branch}, and hence at least two roots per branch. This is because
\begin{equation}
c_8 c_{-8} \geq 0,
\end{equation}
which is satisfied in any orbital combination. This property follows from the explicit
expression for $c_{\pm 8}$ given in \citep{BalMik19}. We consider now that $\mathcal P(z)$
is real-valued, and if $c_8 c_{-8}>0$ then $\mathcal P(0)$ and $\mathcal P(\pm\infty)$ have
the same sign. Hence, there are even number of positive and even number of negative roots,
corresponding to either real or imaginary branch. In case when one (or both) $c_{\pm 8}$
vanishes then there is one or more roots $z=0$ or $w=0$, corresponding to the $\mathcal
H\mathcal H$-case with parallel asymptotes \citep{Baluev05}. These zero roots can be
arbitrarily attributed to either branch, so the number of roots per branch can always be
treated even.

The upper limit on the number of real roots is uncertain even in the $\mathcal E\mathcal E$
case. It cannot exceed $16$, the algebraic degree of $g(u)$, but simulations never revealed
more than $12$ real roots \citep{KholshVas99,BalMik19}. Sometimes we obtained $14$-root and
even $16$-root occurrences using the standard {\sc double} precision, but with {\sc long
double} arithmetic these cases appeared to have no more than $12$ real roots. Moreover, in
the testcase considered below we likely had none numerically reliable cases with $12$ real
roots. Considering hyperbolas, we detected a small fraction of quite reliable $12$-root
occurrences, but when restricted to only $z\geq 0$ (the main branch of $\mathcal O$), we
had only $8$ real roots at most in the $\mathcal H\mathcal E$ case, and $10$ at most in the
$\mathcal H\mathcal H$ case.

Since the number of roots can be highly variable, it is a frequent case when there are
close roots that cannot be easily resolved due to numeric errors. This effect was important
in the $\mathcal E\mathcal E$ case, where it could lead to lost roots of $g(u)$, due to
their misclassification as complex rather than real. This effect lead sometimes to
overestimated distance because some critical points might appear lost. The solution was to
track numeric errors of the roots and to select the real ones based on their estimated
numeric uncertainty (so that some formally complex roots of $g(u)$ with small imaginary
part could be tested together with purely real ones). In this work we expand this same
approach to the hyperbolic orbits as well.

\section{Numerical algorithm}
\label{sec_alg}
The general computing sequence remains the same as in \citep{BalMik19}.
\begin{enumerate}
\item Compute the coefficients of $g(u)$ and their numeric uncertainties by means of the
Discrete Fourier Transform (DFT).

\item Determining reasonable starting approximations for some of the $g(u)$ roots.

\item Find all roots of the polynomial $\mathcal P(z)$ or $\mathcal Q(w)$ in the complex
plane by Newtonian iterations (this task dominates, taking about $60\%$ of computing time).

\item Estimate roots uncertainties, and select those roots that correspond to real orbital
positions (and lying on real hyperbola branches).

\item Among these roots, select the one providing the minimum distance and perform
Newtonian 2D iterations to refine this minimum distance and its position.
\end{enumerate}
Now let us comment each stage of this sequence in our updated algorithm.

\subsection{Determining polynomial coefficients}
The coefficients of $g(u)$ are computed by DFT as if we always dealt with the $\mathcal
E\mathcal E$ case, just replacing the definition of $g$ appropriately. We compute a set of
values $g_k = g(u_k)$ for equally spaced $u_k$, and then apply the DFT to obtain all $c_k$.
We use an excessive number of $u_k$ to also compute a few $c_k$ with $|k|>8$. These
quantities should be zero in theory, so they can be helpful to estimate numeric errors in
$c_k$.

This scheme can be replicated literally in the $\mathcal E\mathcal H$ case, when $g(u)$ is
a trigonometric polynomial with real coefficients. In the $\mathcal H\mathcal E$ and
$\mathcal H\mathcal H$ cases $g(u)$ becomes a hyperbolic polynomial, and $u$ attains the
meaning of the hyperbolic anomaly. In this case we could compute the coefficients of
$\mathcal P(z)$ or $\mathcal Q(w)$ by performing their polynomial interpolation inside a
real segment. This would allow us to avoid complex numbers, as all quantities become
real-valued. However, this infers dealing with large differences (large $z$ and small $z$
present simultaneously), so we suspected that this scheme might appear less numerically
stable. Therefore, we decided to consider $g(u)$ as a trigonometric polynomial of the
imaginary argument (which has the meaning of eccentric anomaly). In this case $g_k$ become
complex quantities, but the DFT should always produce real $c_k$, thanks to the property
$g(iu) = g^*(-iu)$. The latter property can also be used to halve the number of complex
multiplications needed for the DFT. The numeric complexity of this DFT remains the same as
in the $\mathcal E\mathcal E$ case, where we used only real multiplications at this stage.

\subsection{Determining starting approximations for the roots}
As shown in \citep{BalMik19}, smart selection of starting approximations to the roots of
$g(u)$ may significantly improve performance as well as numeric stability of the algorithm.
In this version of the library, we adopt generally the same scheme here, though augmented
in several aspects. This general scheme involves approximations for $4$ complex roots
extracted first, then $4$ guaranteed real roots, and for the remaining roots we adopt the
same scheme as for the first $4$ ones. This allowed us to reduce the cumulative number of
Newtonian iterations, though we need to emphasize that this is merely a \emph{statistical}
effect, appearing when a lot of orbit pairs are processed. Some individual orbit pairs may
reveal a specific behavior due to peculiar properties of the roots.

In the $\mathcal E\mathcal E$ case, we started iterating the first (complex) root from
$z_1=0$. The second root starts from $z_2=1/z_1^*$, third from $z_3=1/z_2^*$ and so on.
Thus we intermittently extract small and large roots, approaching the unit circle. This
scheme was based on the empiric observation that complex roots concentrate near $z=0$ and
$w=0$. This scheme is preserved in the updated algorithm, and extended literally to the
$\mathcal E\mathcal H$ case. However, in the $\mathcal H\mathcal E$ and $\mathcal H\mathcal
H$ cases we use more specific starting approximations based on the investigation of
statistical distribution of the roots in the complex plane (see Fig.~\ref{fig_rpdf}
discussed below). In particular, in the $\mathcal H\mathcal H$ case we start $z_1$ from
$\exp(0.4\pi i)$, $z_2$ is started from $z_1^*$, $z_3$ is started from $+i$, $z_4$ from
$z_3^*$. In the $\mathcal H\mathcal E$ case the value $+i$ is replaced by $\exp(0.6\pi i)$.

Starting approximation for the $4$ guaranteed real roots are based on the assumption that
critical points are often located not far from the orbital nodes. But the hyperbolic orbits
introduce additional difficulties in this concern. In particular, we should take care of
the real/imaginary hyperbola branches. In the $\mathcal H\mathcal H$ case exactly one
starting point should be placed per each branch combination, because we cannot guarantee
that any combination has more than a single critical points. In the $\mathcal H\mathcal E$
and $\mathcal E\mathcal H$ case we should distribute two starting points per hyperbolic
branch. So, depending on the orbital combination, we adopt the following sets of starting
approximations for critical points:
\begin{enumerate}
\item $\mathcal E\mathcal E$: two internodal distances, each lying entirely on either side from
the focus and two distances connecting points separated by $\pm \pi/2$ from the nodes (in
true anomaly), and also lying on the same side from the focus. The first pair usually
corresponds to two local minima, while the second one usually describes saddle points.

\item $\mathcal E\mathcal H$ or $\mathcal H\mathcal E$: if each $\mathcal H$-branch passes
through just a single node then, per each branch, select one internodal distance lying
entirely on the same side from the focus (usually minimum), and another internodal distance
containing the focus (usually saddle point).

\item $\mathcal H\mathcal H$: select all $4$ possible internodal distances, if they all
refer to different branch combinations (that is, if each hyperbolic branch passes through
just a single orbital node).

\item The conditions above may be violated if one or both hyperbolic orbits pass through
both its nodes simultaneously, while the corresponding imaginary branch does not intersect
the other orbital plane at all. In this case half of starting approximations should
necessarily involve this imaginary branch. Its orbital position is selected relatively
arbitrarily, through a half-sum of hyperbolic anomalies corresponding to the two nodes on
the real branch.
\end{enumerate}
Approximations of this ``internodal type'' should be good for highly inclined orbit pairs,
and they revealed good efficiency in the testcase from \citep{BalMik19}. However, it
appeared that we did not properly transform angular elements to radians in that testcase
(see Corrigendum to \citealt{BalMik19}). This does not mean that initial test results were
wrong, as they just corresponded to different testcase. But in the corrected one such initial
approximations appeared less efficient and did not reduce the number of Newtonian
iterations as desired. Obviously, this appeared because of typically small orbital
inclination in the asteroid belt. Also, it appeared that such starting solution are less
efficient for hyperbolic orbits as well. We solved the issue by pre-refining every starting
approximation through a single 2D Newtonian iteration of $\rho(u,u')$.

One might argue that this is no better than to make more Newtonian iterations on $\mathcal
P(z)$, but in actuality $\mathcal P(z)$ is less efficient to iterate, because it typically
contains close roots. Such an effect appears because each nodal $u$ can be paired with
either of two nodal $u'$. That is, internodal critical points of $\rho(u,u')$ usually
project to two close pairs when considered in terms of just $u$ or just $u'$. But when
dealing with $\rho(u,u')$ they are separated well, thus avoiding the effect of close roots.
Hence, even a single 2D iteration allowed to determine a good starting approximation for
$u$. In turn, this approach allowed us to save $\sim 20-30$ Newtonian iteration of
$\mathcal P(z)$ by the cost of making just $4$ additional 2D iterations of $\rho(u,u')$.

The rest of the roots is determined automatically: starting from near-zero $z$ or near-zero
$w$ in case of complex $c_k$ ($\mathcal E\mathcal E$ or $\mathcal E\mathcal H$), or from a
random point on unit circle for real $c_k$ ($\mathcal H\mathcal E$ or $\mathcal H\mathcal
H$). The latter choice was motivated by the observation that real roots are often more
difficult to locate without a good starting approximations.

\subsection{Finding all complex roots}
The roots of $\mathcal P(z)$ are found one by one, by means of running the Newtonian root
search with subsequent division of $\mathcal P(z)$ by the corresponding linear factor
$z-z_k$. we introduced a few minor changes to this scheme.

First, we extract the roots until a quartic polynomial is obtained, rather then a quadratic
one. The last four roots of the quartic polynomial are computed using explicit formulae of
the Ferrari method. Since these formulae are more complicated and may have a reduced
numeric stability sometimes, we also refine each of these $4$ roots by a single Newtonian
iteration applied to that quartic polynomial. Solving the quartic polynomial in such a way
appeared a bit faster.

Second, to further improve numeric stability, we apply Newton iterations to either
$\mathcal P(z)$, if $|z|<1$, or $\mathcal Q(w)$, if $|z|>1$. In the previous version of the
algorithm we rejected this approach because it frequently lead to looping conditions, when
we infinitely jump between the same points. Now we avoided this issue by performing the
mode change in a more ``lazy'' manner. Namely, we keep the $|z|<1$ mode until $|z|^2$
reaches $5$, and the mode $|z|>1$ is kept while $|z|^2>1/2$.

Third, while in the elliptic case the paired complex roots obey the rule $z_2 = 1/z_1^* =
z_1/|z_1|^2$, in case when the first orbit is hyperbolic it should be replaced by $z_2 =
z_1^*$.

\subsection{Real roots selection}
In the $\mathcal E\mathcal E$ and $\mathcal E\mathcal H$ cases $g(u)$ is a trigonometric
polynomial, so we should select the roots lying on the unit circle $|z|=|w|=1$. We may then
use the same criterion as in \citep{BalMik19}:
\begin{equation}
\Delta_z = \frac{\left|\log |z| \right|}{\nu \varepsilon_z} \leq 3,
\label{thrt}
\end{equation}
where $\nu$ is a manual normalizing factor and $\varepsilon_z$ is the relative root
uncertainty (estimated literally as in \citealt{BalMik19}). This criterion can be rewritten
as
\begin{equation}
\left|\Re \log z \right| \leq 3\nu \varepsilon_z.
\end{equation}

Now we have $\mathcal H\mathcal E$ and $\mathcal H\mathcal H$ cases, in which $g(u)$ is a
hyperbolic polynomial. The coefficients of $\mathcal P(z)$ are real, and we are interested
in only real positive $z$ (that correspond to real $u$). In this case the criterion becomes
\begin{equation}
\left|\Im \log z \right| \leq 3\nu \varepsilon_z,
\end{equation}
so in place of~(\ref{thrt}) we have
\begin{equation}
\Delta_z = \frac{\left|\arg z \right|}{\nu \varepsilon_z} \leq 3.
\label{thrh}
\end{equation}
Notice that this criterion also filters away all roots corresponding to the imaginary
branch of $\mathcal H$ (negative $z$). Still, in $\mathcal H\mathcal H$ case it might occur
that some roots with $z>0$ correspond to the imaginary branch of the second orbit $\mathcal
H'$. We may then obtain negative $\cosh u'$.

\subsection{Final MOID refining and testing the reliability}
Each selected root $z_k$ implies the orbital position as $u_k = \arg z_k$, or $u_k = -\log
z_k$. The orbital position on $\mathcal O'$ is determined using formulae from
\citep{Baluev05}. After that, the both orbital positions may still involve increased
numeric errors and should be refined using 2D Newtonian iterations. The scheme here remains
entirely the same as in \citep{BalMik19}, with obvious replacements for the hyperbolic
motion. The formulae determining numeric uncertainty of the result are also nearly the
same, with the only difference that $\rho(u,u')$ and the product $aa'$ are negative in the
mixed $\mathcal E\mathcal H/\mathcal H\mathcal E$ cases. They all should be replaced by
absolute values in that formulae.

Finally, we apply the following sequence of checks to verify the reliability of the
results:
\begin{enumerate}
\item All roots that passed~(\ref{thrt}) or~(\ref{thrh}) must comply with the requested
least accuracy: $\nu\varepsilon_z<\delta_{\max}$.

\item The minimum of $\Delta_z$ among all the roots that failed~(\ref{thrt})
or~(\ref{thrh}) must exceed $10$, meaning that there is no other suspicious root
candidates. That is, the families of selected real and other roots must be separated by a
clear gap.

\item The number of roots that passed~(\ref{thrt}) or~(\ref{thrh}) must be even and no
smaller than $4$ (for $\mathcal E\mathcal E$ and $\mathcal E\mathcal H$ cases), or even and
no smaller than $2$ (for $\mathcal H\mathcal E$ and $\mathcal H\mathcal H$ cases).

\item After the 2D refining, the Hessian $\mathbmss H_{\rm rsd}$ of $\rho(u,u')$ is
positive-definite (for $\mathcal E\mathcal E$ and $\mathcal H\mathcal H$ cases), or
negative-definite (for $\mathcal E\mathcal H$ and $\mathcal H\mathcal E$ cases).

\item On the 2D refining stage, the total cumulative change in $u$ satisfies the condition
$|\Delta u| < \delta_{\max}$.
\end{enumerate}
They mostly replicate those from \citep{BalMik19}, with minor corrections.

\subsection{Extending the fallback algorithm to hyperbolas}
In addition to the basic fast method based on $g(u)$ root-finding, the {\sc distlink}
library implemented a brute force-like minimization of the distance function $\rho(u,u')$
with respect to $u$ (the other position $u'$ is determined through $u$ analytically). This
more slow method can be used as an additional alternative when the basic one signals a
warning. In this work we generalized this method to the hyperbolic orbits as well. The
general scheme remains entirely the same as in \citep{BalMik19}. The changes are only due
to the scan range for $u$ which is determined by new formulae. Its method was based on an
observation that MOID is usually located close to the orbital nodes, and this idea is
further developed in \ref{sec_hrange} for orbit pairs involving hyperbolas. As in the
elliptic case, the algorithm may automatically swap the orbits to scan a smaller range in
$u$.

\section{Practical validation and benchmarks}
\label{sec_perf}
We tested our algorithm on the first $10000$ numbered asteroids from the Lowell observatory
catalogue \texttt{astorb.dat}\footnote{See url ftp://ftp.lowell.edu/pub/elgb/astorb.html}.
This implies $\sim 10^8$ orbit pairs. We used exactly the same version of the catalogue as
in \citep{BalMik19}, but here we correctly transform angular elements (see Corrigendum to
that work).

To obtain hyperbolic orbits, we transformed the asteroid orbital eccentricities as
$e\mapsto 1-\log e$, changing the sign of semimajor axis. In this way we generated $\sim
10^8$ orbit pairs for each of our orbital configurations: $\mathcal E\mathcal E$, $\mathcal
E\mathcal H$, $\mathcal H\mathcal E$, and $\mathcal H\mathcal H$. Notice that computation
is asymmetric in its nature both for our algorithm and for the \citet{Gronchi05} code, so
the result may appear different for ${\rm MOID}(\mathcal O,\mathcal O')$ for ${\rm
MOID}(\mathcal O',\mathcal O)$. Normally, this should be a small difference due to only
numeric errors, but it may appear large if we faced some serious algorithm failure like
missing critical point.

In this work we used the Intel Core i7 configuration explained in \citep{BalMik19} and the
standard {\sc double} floating-point arithmetic (unless otherwise stated).

\begin{table}
\caption{Frequency of potentially unreliable occurrences.}
\begin{tabular}{lll}
\hline
case & single ${\rm MOID}(\mathcal O,\mathcal O')$ & ${\rm MOID}(\mathcal O,\mathcal O')$ and\\
& & ${\rm MOID}(\mathcal O',\mathcal O)$ \\
\hline
$\mathcal E\mathcal E$ & one per $14600$  & one per $413000$ \\
$\mathcal E\mathcal H$ & one per $66000$  & one per $10^8$   \\
$\mathcal H\mathcal E$ & one per $348000$ & ---~as~above~--- \\
$\mathcal H\mathcal H$ & one per $2580$   & one per $55900$  \\
\hline
\end{tabular}
\label{tab_freq}
\end{table}

The frequency of unreliable cases that failed some of the post-computing requirements
above, are given in Table~\ref{tab_freq}. Notice that these frequencies in the $\mathcal
E\mathcal E$ case appeared somewhat higher than mentioned in \citep{BalMik19} because the
new values refer to the corrected test case which has more near-coplanar occurrences. The
number of potentially unreliable computations is increased in the $\mathcal H\mathcal H$
case because it frequently infers that MOID segment is located rather far from the focus,
basically between the asymptotes as skew lines. In the degenerated case when some
asymptotes are parallel, the MOID can be even achieved at the infinity \citep{Baluev05}. In
such cases, additional numeric difficulties appear because of subtraction of large close
vectors when computing the MOID. This results in additional numeric errors causing the
violation of our post-condition that error in $u$ should be below the maximum allowed
error. This effect seems difficult to avoid, but from the other side such violation appears
rather formal, because it is unrelated to possible lost roots.

Among occurrences that were reported as unreliable in either ${\rm MOID}(\mathcal O,
\mathcal O')$ or ${\rm MOID}(\mathcal O', \mathcal O)$ run, nearly all appeared to have
correct MOID in turn. In the $\mathcal E\mathcal E$ run we detected only $11$ occurrences
when either of the two runs gave wrong distance. However, in all that cases the larger
distance was always flagged properly as unreliable, while the other one appeared in
agreement with the MOID provided by \citet{Gronchi05} code. In all the cases when both
computations were flagged as unreliable, both MOIDs actually appeared correct. There were
only $5$ occurrences with wrong distance in the $\mathcal H\mathcal E/\mathcal E\mathcal H$
case, and $4$ ones in the $\mathcal H\mathcal H$ one. All such events were properly flagged
as well.

\begin{figure*}[!t]
\begin{tabular}{@{}c@{}c@{}}
\includegraphics[width=0.49\textwidth]{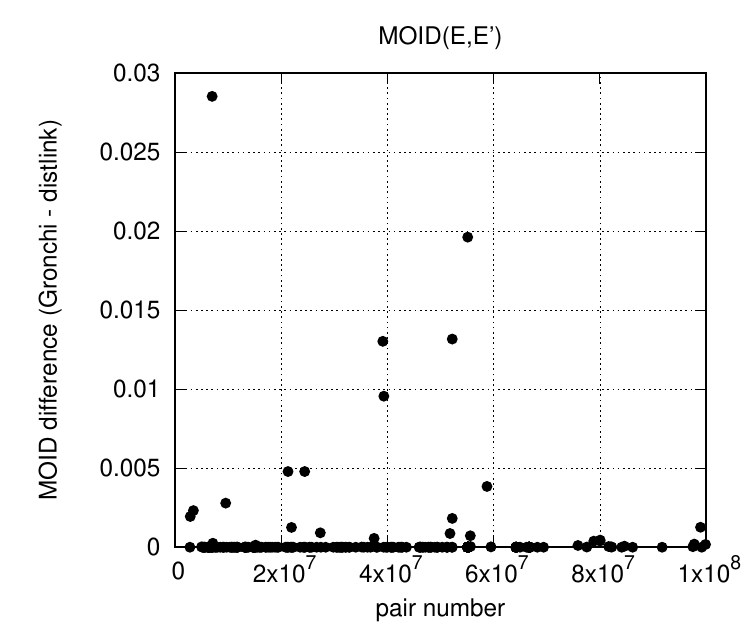} &
\includegraphics[width=0.49\textwidth]{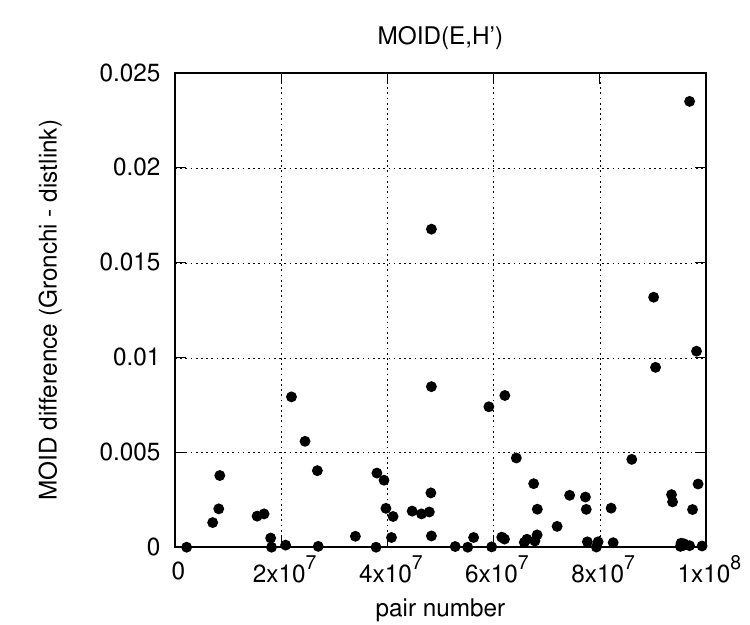}\\
\includegraphics[width=0.49\textwidth]{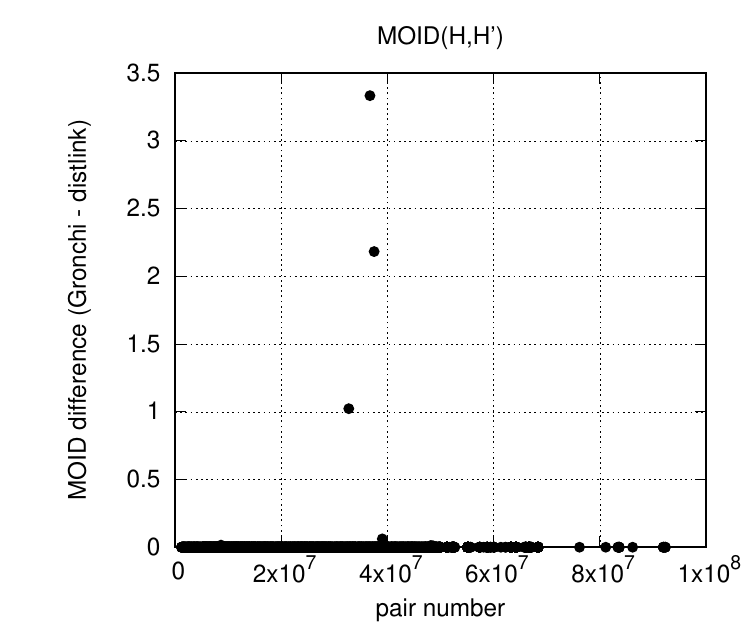} &
\includegraphics[width=0.49\textwidth]{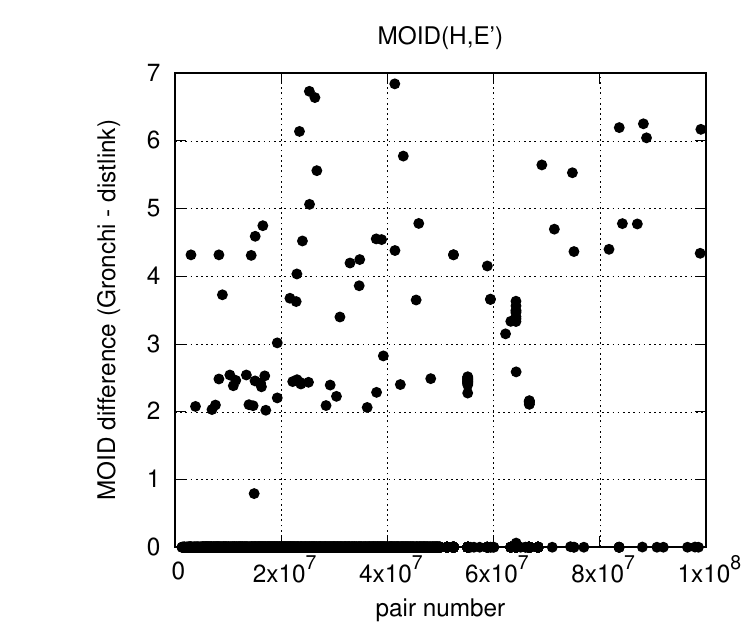}\\
\end{tabular}
\caption{The difference between MOID values computed by the Gronchi's code and by our
algorithm (labelled as {\sc distlink}). Panels refer to different orbital combinations as
labelled. All differences smaller than $10^{-11}$~AU (in absolute value) were removed. No
points below zero were detected in these conditions, meaning that our code always returned
smaller value than the Gronchi's one.}
\label{fig_Gtest}
\end{figure*}

We provide a comparison of our algorithm with the Gronchi code in Fig.~\ref{fig_Gtest}. We
compute the differences of the Gronchi code MOID minus the MOID obtained by our code. If
either algorithm self-diagnosed a warning/failure, we removed this occurence from the
consideration. We may see that in every orbital combination there are multiple occurrences
when Gronchi code obtained clearly overestimated MOID value (i.e., it missed the true
global minimum). We did not detect opposite occurrences, possibly except for those with
difference below $10^{-11}$~AU (they are simply the effect of round-off errors).
Summarizing, we did not find an occurence in which the {\sc distlink} would yield clearly
wrong MOID value without setting the unreliability flag.

\begin{figure}[!t]
\includegraphics[width=\linewidth]{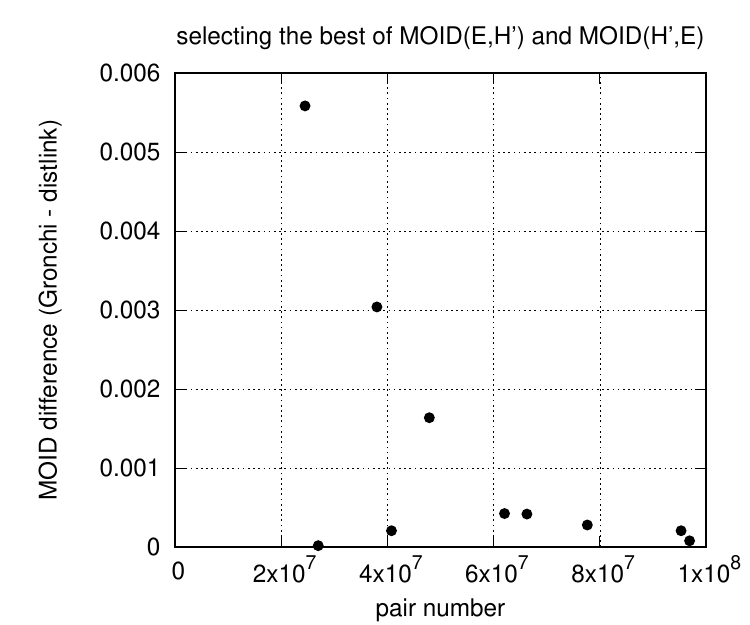}
\caption{The difference between MOID values computed by the Gronchi's code and by our
algorithm (labelled as {\sc distlink}). Similar to Fig.~\ref{fig_Gtest}, but here we
compute the MOID twice by swapping the orbits and select the ``best'' result (see text).
Only the mixed $\mathcal H\mathcal E/\mathcal E\mathcal H$ case is shown, because there
were none significant differences in the other two combinations.}
\label{fig_Gtestswap}
\end{figure}

We also considered another test setting when we compute the MOID twice by swapping the
orbits. The {\sc distlink} algorithm may be called second time only if the first
computation signaled a warning, so this does not essentially slow down the cumulative
performance. The Gronchi code, however, does not seem to diagnose its failures well, as we
have demonstrated above. Therefore, we always run it twice in this test, roughly doubling
the computing time. Still, there are orbit pairs that could not be processed well by
Gronchi code even with such swapping, while our code yielded significantly smaller MOID on
them (see Fig.~\ref{fig_Gtestswap}). We found such pairs only in the mixed $\mathcal
E\mathcal H/\mathcal H\mathcal E$ case, while the $\mathcal E\mathcal E$ and $\mathcal
H\mathcal H$ cases demonstrated good agreement between {\sc distlink} and Gronchi code.
However, such pairs were found the initial (uncorrected) testcase from \citep{BalMik19} in
the $\mathcal E\mathcal E$ case.

\begin{figure*}[!t]
\begin{tabular}{@{}c@{}c@{}}
\includegraphics[width=0.49\textwidth]{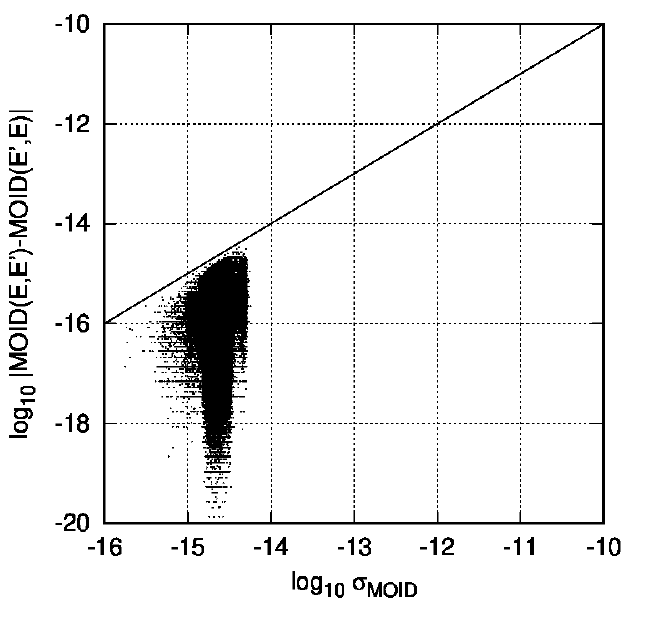} &
\includegraphics[width=0.49\textwidth]{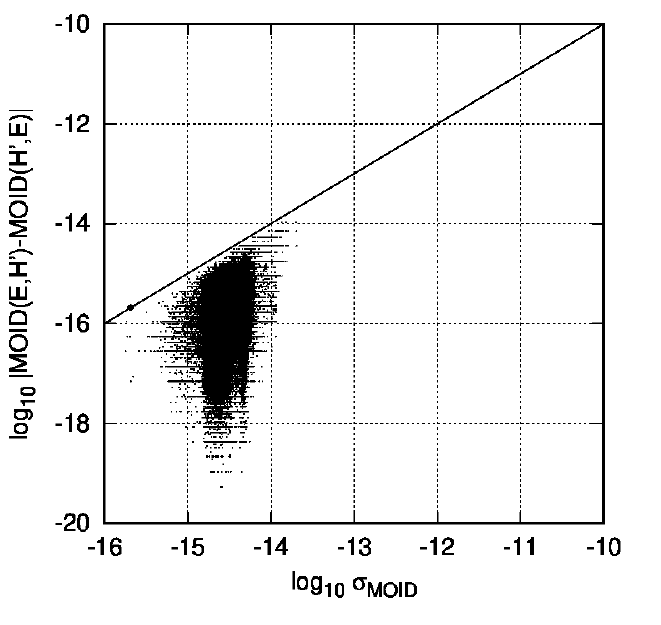}\\
\includegraphics[width=0.49\textwidth]{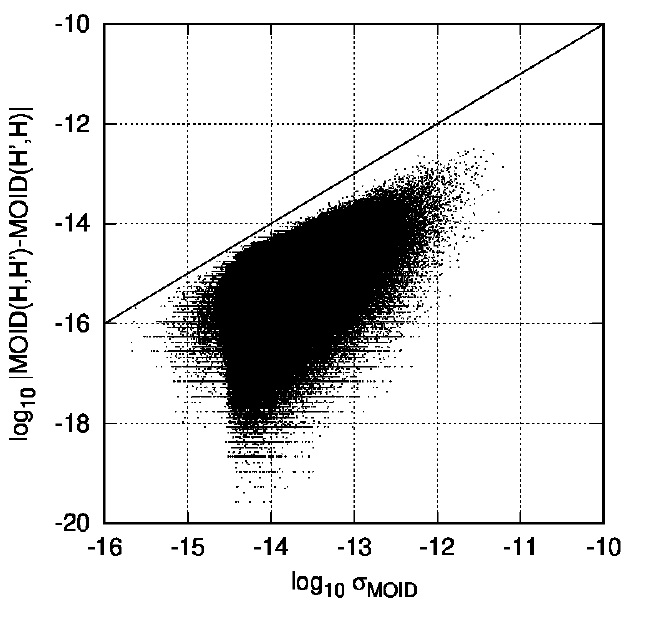} &
\includegraphics[width=0.49\textwidth]{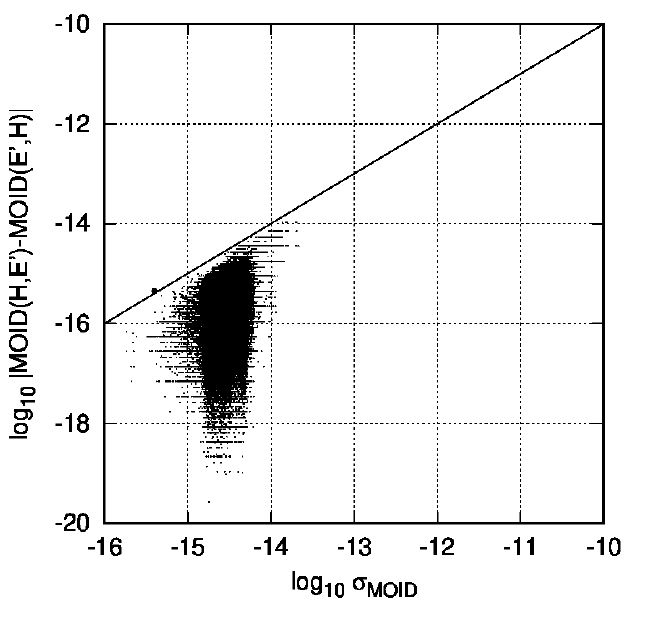}\\
\end{tabular}
\caption{Distribution of the estimated uncertainties
$\sigma_\mathrm{MOID}$ versus an empiric error measure $\left|\mathrm{MOID}(\mathcal O,
\mathcal O') - \mathrm{MOID}(\mathcal O', \mathcal O)\right|$. Panels refer to different
orbital combinations as labelled ($\mathcal H\mathcal E$ and $\mathcal E\mathcal H$ cases
should be statistically equivalent here).}
\label{fig_testerr}
\end{figure*}

In Fig.~\ref{fig_testerr} we compare the quadrature sum of the reported MOID uncertainties,
$\sigma_\mathrm{MOID}=\sqrt{\sigma_\mathrm{MOID(\mathcal E,\mathcal E')}^2 +
\sigma_\mathrm{MOID(\mathcal E',\mathcal E)}^2}$, with the difference
$|\mathrm{MOID}(\mathcal E,\mathcal E')-\mathrm{MOID}(\mathcal E',\mathcal E)|$ that can be
deemed as an empiric estimate of the actual MOID error. We may conclude that our algorithm
provides rather safe and realistic assessment of numeric errors, intentionally somewhat
pessimistic. We found just two occurrences when the empiric error slightly exceeded the
predicted uncertainty, but this violation appears very formal and insignificant. Notice
that initially we did not expect to put a strict error limit, this behaviour appears partly
curious.

\begin{figure*}[!t]
\begin{tabular}{@{}c@{}c@{}c@{}}
\includegraphics[width=0.33\textwidth]{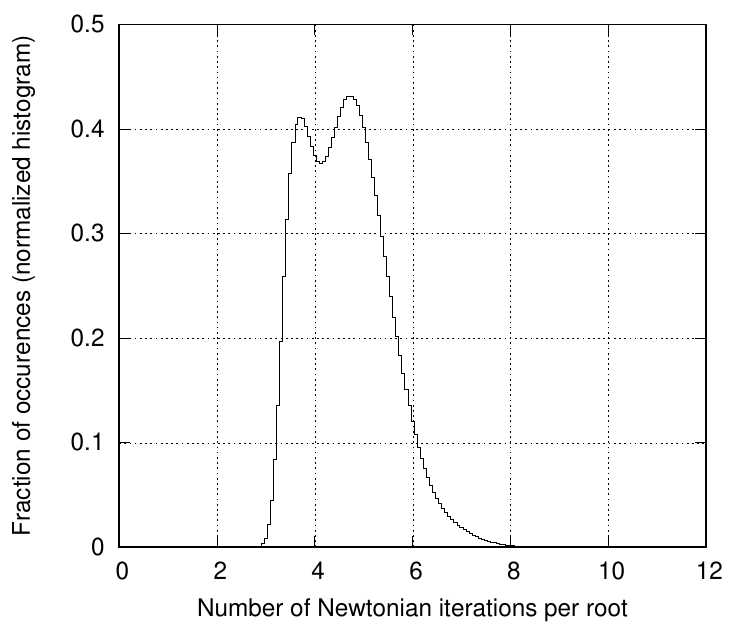} &
\includegraphics[width=0.33\textwidth]{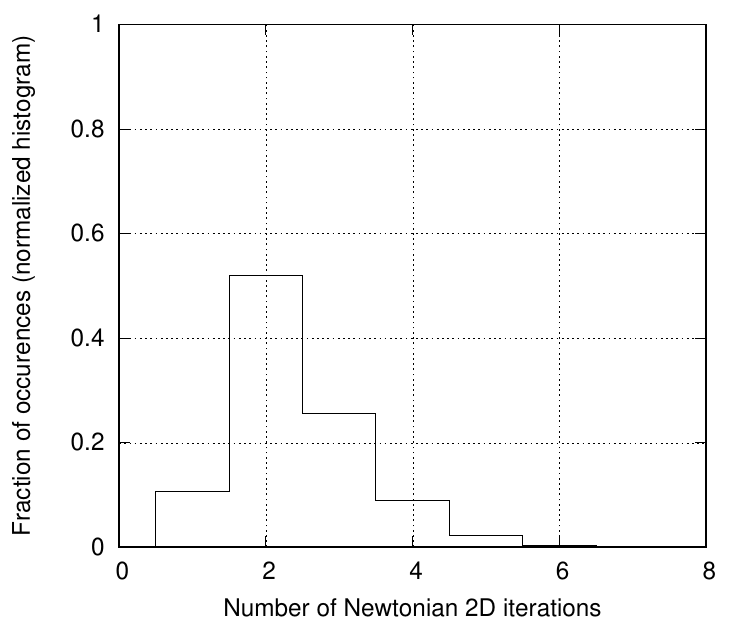} &
\includegraphics[width=0.33\textwidth]{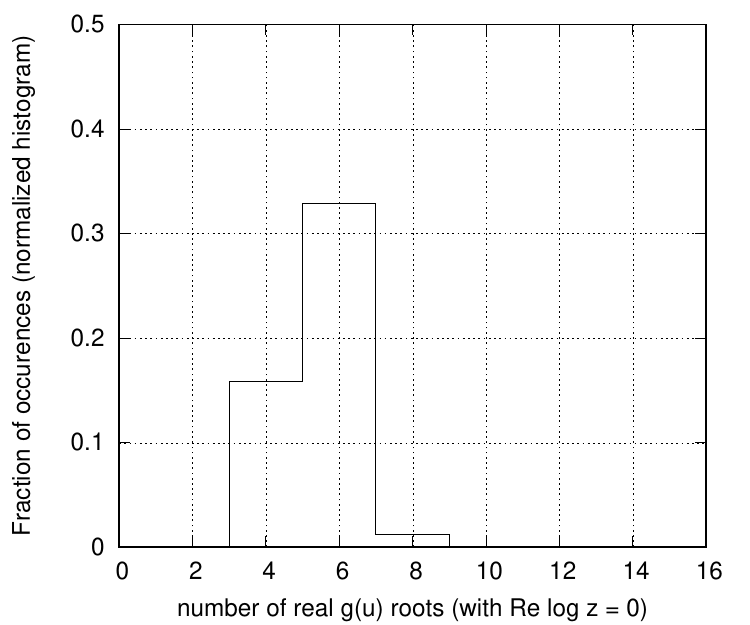} \\
\includegraphics[width=0.33\textwidth]{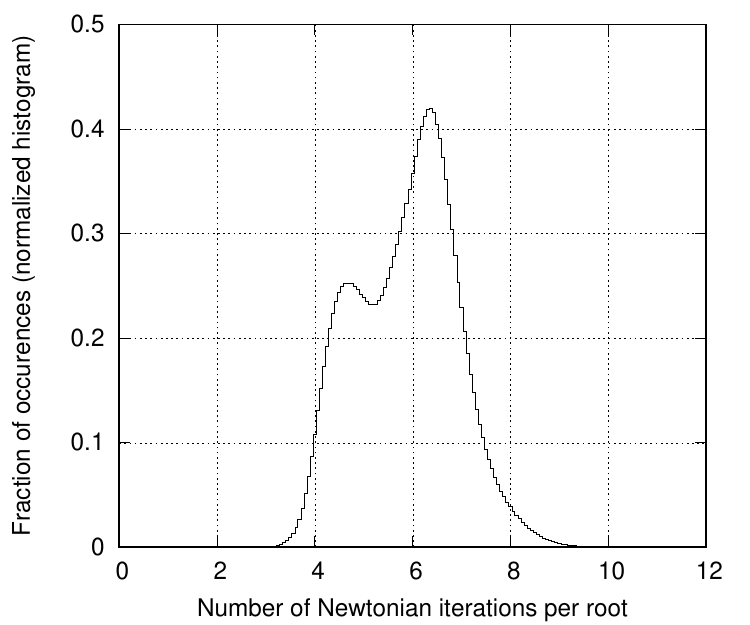} &
\includegraphics[width=0.33\textwidth]{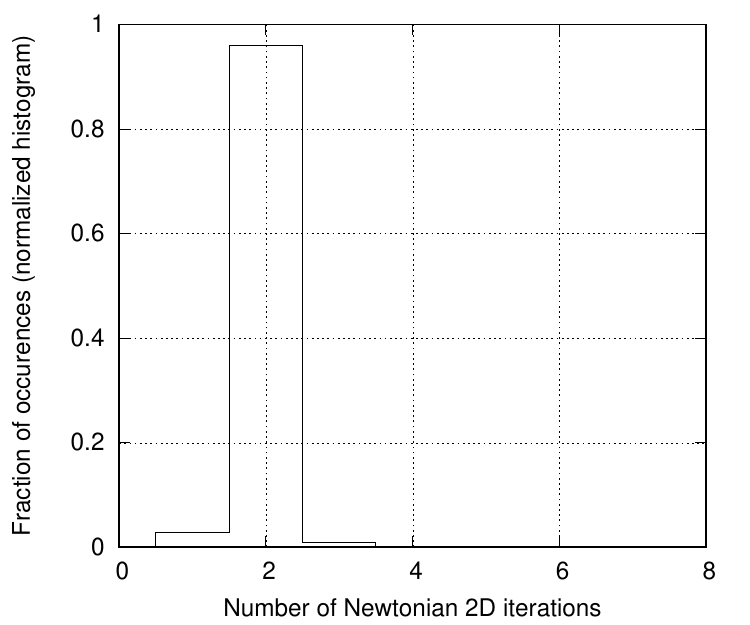} &
\includegraphics[width=0.33\textwidth]{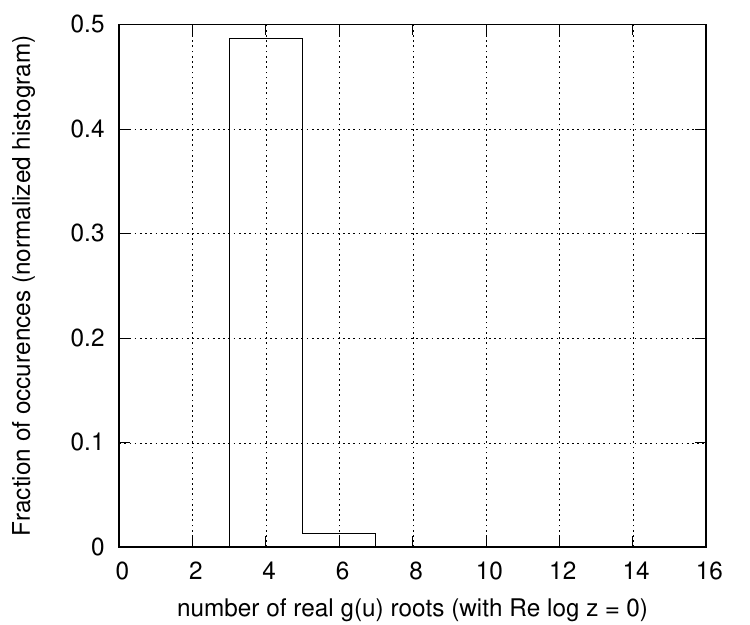} \\
\includegraphics[width=0.33\textwidth]{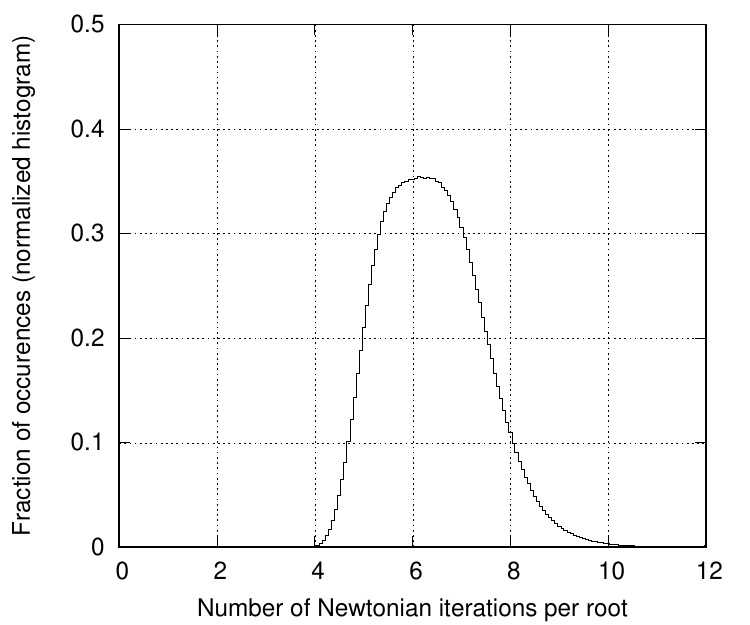} &
\includegraphics[width=0.33\textwidth]{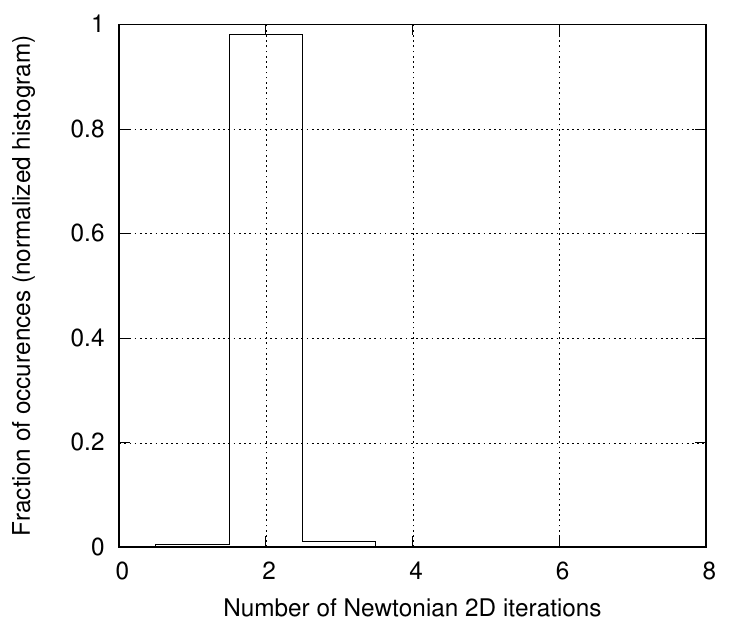} &
\includegraphics[width=0.33\textwidth]{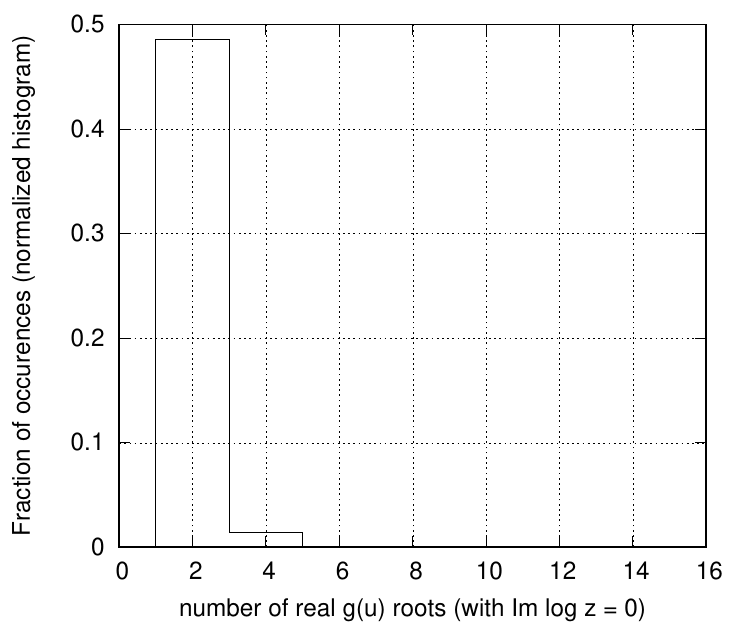} \\
\includegraphics[width=0.33\textwidth]{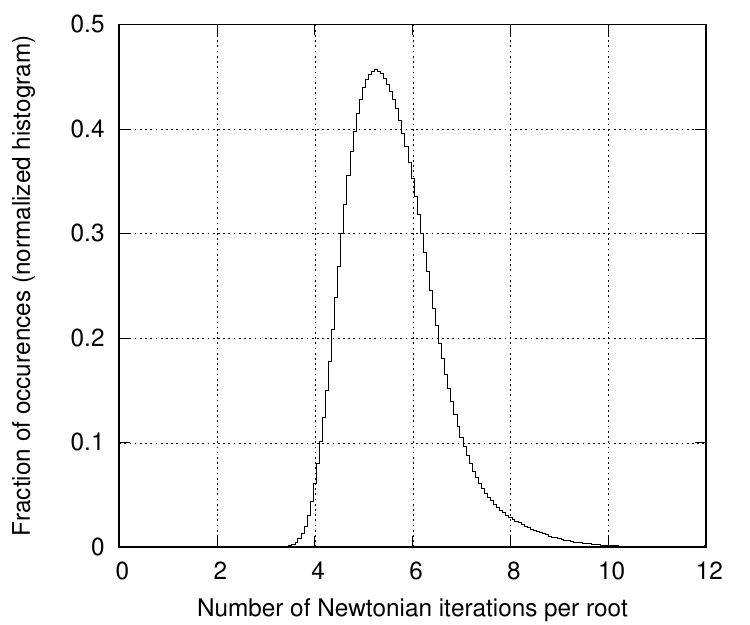} &
\includegraphics[width=0.33\textwidth]{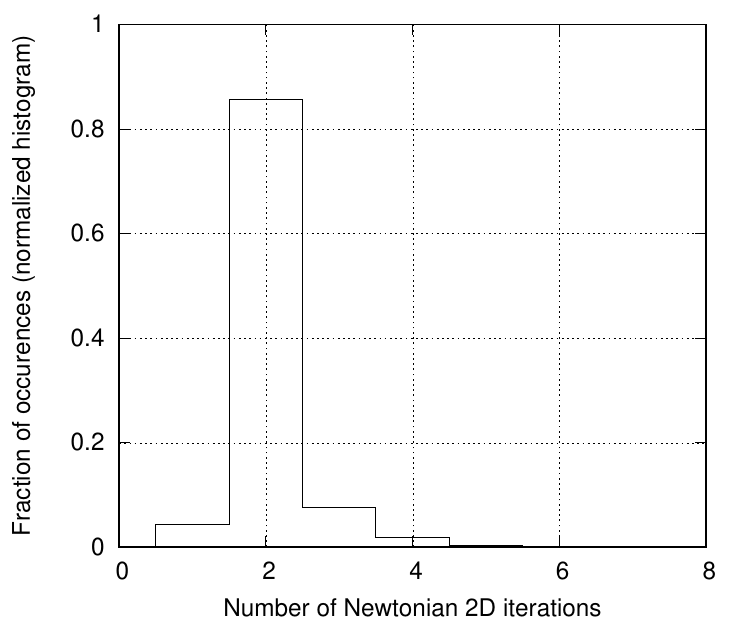} &
\includegraphics[width=0.33\textwidth]{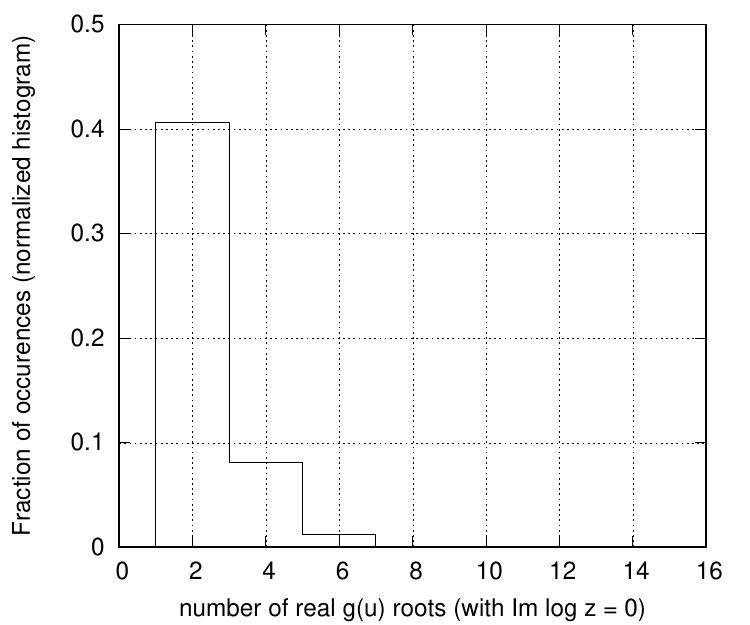} \\
\end{tabular}
\caption{Histograms for the number of Newtonian iterations spent per root of $g(u)$ (left), of 2D
Newtonian iterations on the refine stage (middle), and for the number of real roots. The
four rows refer to the $\mathcal E\mathcal E$, $\mathcal E\mathcal H$, $\mathcal H\mathcal
E$, $\mathcal H\mathcal H$ cases (top to bottom). The histograms were normalized by the bin
width to render the probability density function for the quantity labelled in the abscissa.}
\label{fig_testiter}
\end{figure*}

In Fig.~\ref{fig_testiter} we provide additional statistical distributions related to our
algorithm: the number of Newtonian iterations of $\mathcal P(z)$, of 2D Newtonian
iterations, and of the total number of real roots of $g(u)$.

\begin{figure*}[!t]
\begin{tabular}{@{}c@{}c@{}}
\includegraphics[width=0.49\textwidth]{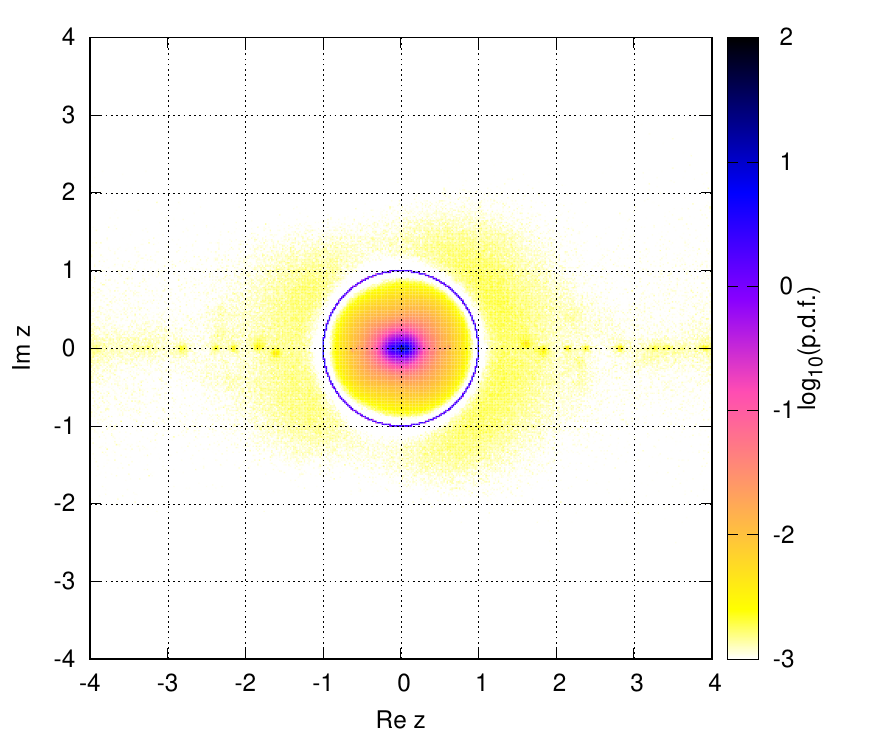} &
\includegraphics[width=0.49\textwidth]{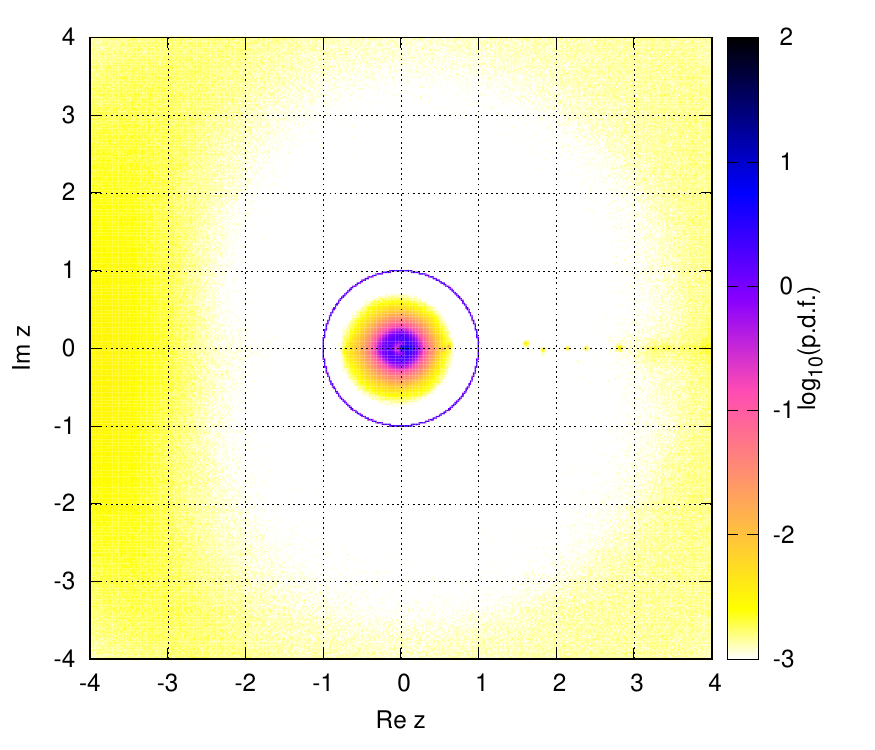}\\
\includegraphics[width=0.49\textwidth]{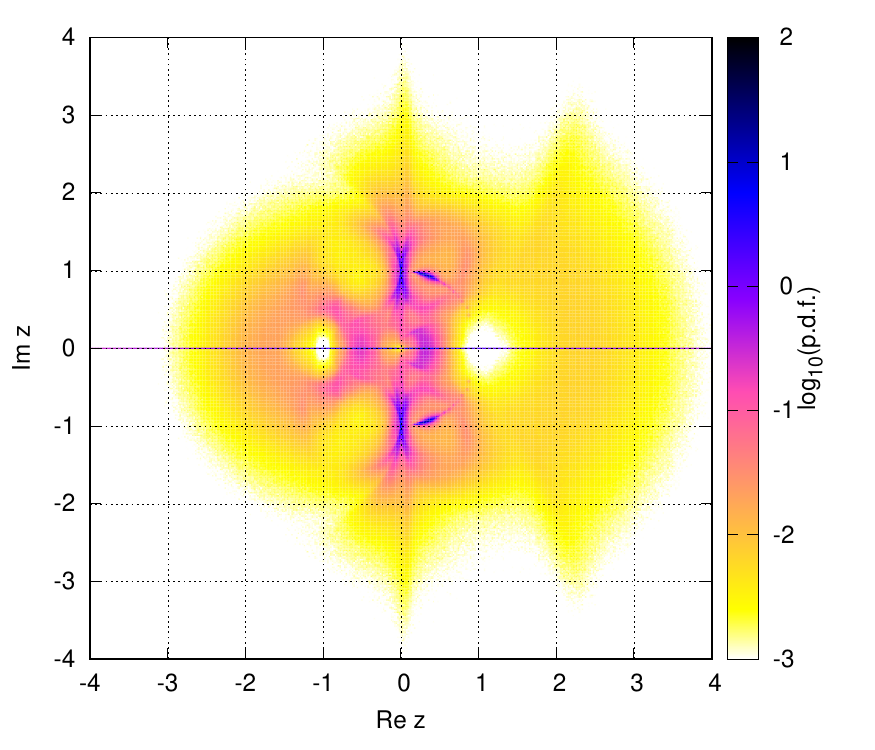} &
\includegraphics[width=0.49\textwidth]{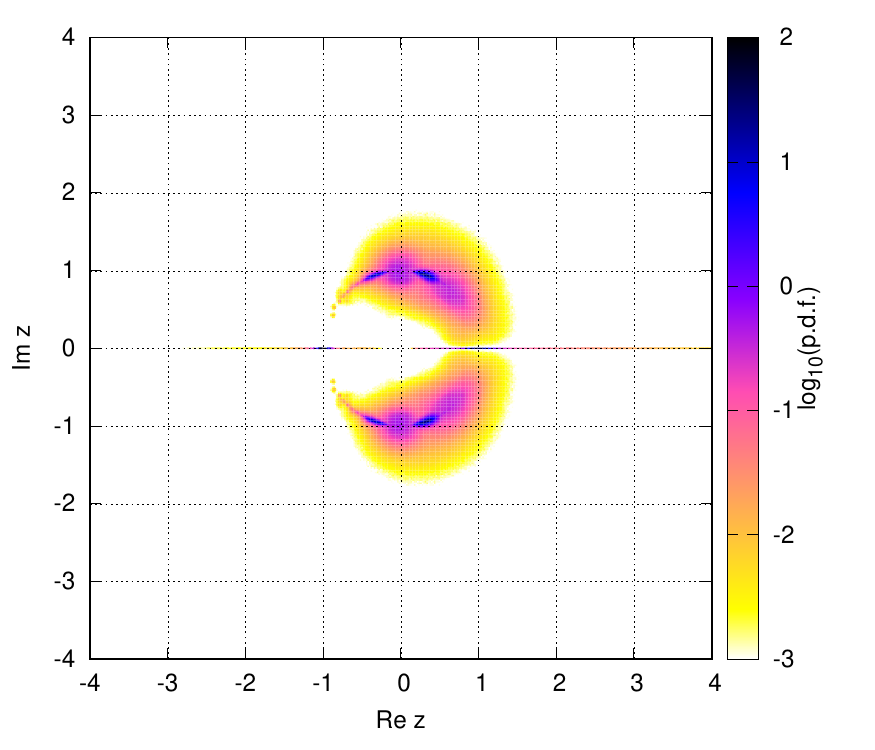}\\
\end{tabular}
\caption{Distribution of $g(u)$ roots in the complex $z$-plane, based on our primary
testcase. Panels show the $\mathcal E\mathcal E$ (top-left), the $\mathcal E\mathcal H$
(top-right), the $\mathcal H\mathcal E$ (bottom-right), and the $\mathcal H\mathcal H$
(bottom-left) cases. The color reflects the level of normalized 2D histogram that estimates
the corresponding probability density function (p.d.f.)}
\label{fig_rpdf}
\end{figure*}

Finally, we investigated the distributions of the roots in the complex plane (in terms of
$z$ variable). They are plotted in Fig.~\ref{fig_rpdf} in the form of 2D histograms, for
all $4$ orbital combinations. We can see that whenever $g(u)$ is trigonometric ($\mathcal
E\mathcal E$ and $\mathcal E\mathcal H$), complex roots concentrate near $z=0$ (and,
concequently, near $w=0$ or $z=\infty$). The real (in terms of $u$) roots are well
separated from the complex ones, there is rather small density near the unit circle (except
for this circle itself). As expected, these two distributions are nearly radially
symmetric, the roots do not reveal remarkable concentrate in any angular directions.

The cases with hyperbolic $g(u)$ appear more intriguing. In this case real (in terms of
$u$) roots lie on the real positive ray $z>0$, but we can see quite complicated and
asymmetric structure in the complex domain, especially detailed near the unit circle. In
the $\mathcal H\mathcal H$ case we can see $4$ obvious concentrations: two at $z=\pm i$
and two with $|z|=1$ and $\arg z \approx \pm 0.4\pi$, while $z=\pm 1$ are avoided (except
for purely real roots). In the $\mathcal H\mathcal E$ case, the $z=\pm i$ concentrations
are replaced by those with $|z|=1$ and $\arg z\approx \pm 0.6\pi$. It is unclear, how much
this puzzling structure is sensitive to our particular testcase. The primary narrow
concentrations are difficult to explain by statistical properties of test orbits.

\begin{table*}
\caption{Performance tests on the first $10000$ asteroids from the Main Belt: average CPU time per MOID.}
\begin{center}
\begin{tabular}{lcccc}
\hline
case                   & \multicolumn{2}{c}{{\sc double} arithmetic} & \multicolumn{2}{c}{{\sc long double} arithmetic} \\
                       & {\sc distlink}& Gronchi code  & {\sc distlink}& Gronchi code   \\
                       & (fast alg.)   &               & (fast alg.)   &                \\
\hline
$\mathcal E\mathcal E$ & $23$~$\mu$s   & $33$~$\mu$s   &  $69$~$\mu$s  &  NA   \\
$\mathcal E\mathcal H$ & $22$~$\mu$s   & $28$~$\mu$s   &  $76$~$\mu$s  &  NA   \\
$\mathcal H\mathcal E$ & $30$~$\mu$s   & $32$~$\mu$s   &  $84$~$\mu$s  &  NA   \\
$\mathcal H\mathcal H$ & $29$~$\mu$s   & $31$~$\mu$s   &  $76$~$\mu$s  &  NA   \\
\hline
\end{tabular}
\end{center}
\label{tab_bench}
\end{table*}

In Table~\ref{tab_bench}, we present our performance benchmarks for this test application.
Here we used either 80-bit {\sc long double} floating-point arithmetic or the 64-bit {\sc
double} one, with all details the same as in \citep{BalMik19}. The code was compiled with
{\sc GCC} and optimized for the local CPU architecture using the same flags as well
(\texttt{-O3 -march=native -mfpmath=sse}). However, the OS was upgraded since that time,
including the {\sc GCC} and {\sc glibc} packages. This might cause a subtle increase of the
performance, $1$~$\mu$s for the Gronchi code (and possibly similar for the {\sc distlink}).

We conclude that our algorithm remains quite competitive in terms of speed. It outperformes
the Gronchi code, though rather marginally if the first orbit is hyperbolic. For the both
algorithms, the $\mathcal H\mathcal E$ configurations appears more slow than $\mathcal
E\mathcal H$, so the latter one is preferred in practice. The exact reason why there is a
remarkable performance drop in our code on a hyperbolic first orbit is still unclear,
perhaps there is still some room to increase the speed in such configurations. The minor
algorithm improvements listed in Sect.~\ref{sec_alg} had only a negligible effect in case
of the {\sc double} arithmetic, but their effect is more remarkable in the {\sc long
double} framework, improving the performance by $10$--$15\%$.

\section{Conclusions}
In the end we would like to discuss further paths of developing the {\sc distlink}
algorithm. We did not yet consider parabolic orbits and other degenerate cases. However,
there are many comets moving on near-parabolic, and even if their eccentricity is not
precisely unit, it can be approximated by unit at some early stage. From the analytic
theory, if either $e=1$ or $e'=1$ the algebraic degree of the associated polynomial is
reduced to $12$, and whenever both the orbits are parabolic the algebraic degree becomes
$9$ \citep{Baluev05}. This would imply certain simplifications and likely significant
speedup of the computation. However, this also requires a more specific treatment of such
cases due to the degeneracy.

Some performance increase to the basic algorithm is still possible. It spends significant
time on finding complex roots of $g(u)$, and therefore one may try to eliminate this
unnecessary work at least partly. This is unfortunately not so easy because we cannot
control the order in which the roots are extracted. Though we made some efforts to extract
$4$ guaranteed real roots in the beginning, this optimization is of a statistical type,
i.e. it works in average but frequently fails for individual configurations. However, to
increase the average performance one may simply stop extracting the roots if the polynomial
has none real roots.

In case when $g(u)$ is hyperbolic with real coefficients $c_k$, this can be verified by the
Descartes' rule of signs, that is, if the coefficients $c_k$ are all positive or all
negative then there are no positive real roots. In case of trigonometric $g(u)$, when $c_k$
are complex, one need to verify the existence of roots with $|z|=1$. This can be done based
on the available classic upper and lower bounds on the polynomial roots; for example, if at
some point $|z_k|<z_{\max}$ and $z_{\max}<1$ then there is no roots with $|z|=1$, hence no
real roots remain in $g(u)$. However, approaches of this type require to track
uncertainties in each $c_k$ which are accumulated after every root extraction (division of
$\mathcal P(z)$ by $z-z_k$). This may result in a loss of numeric reliability due to
unavoidable assumptions about errors, so implementing this method would require a detailed
investigation of possible side effects.

Another way to increase the performance is optimization of complex-valued arithmetics. It
is currently based on the build-in C++ library (through class \texttt{complex}). This
implementation appears numerically reliable (at least for {\sc GCC}), however it includes
significant overheads due to additional checks related to correct support of NaN/Inf
arithmetic and over/underflows. We noticed a roughly double speedup whenever these extra
checks are turned off (the {\sc GCC} flag \texttt{-ffast-math}). The complex multiplication
is then reduced to the school grammar formula $z w = (\Re z \Re w - \Im z \Im w) + i (\Re z
\Im w + \Im z \Re w)$. In this way our library unfortunately generated frequent wrong
results without warnings, indicating its sensitivity to correct handling of such subtle
arithmetic issues. Nevertheless, a remarkable performance increase can be achieved if the
causes of such sensitivity are located in the code and eliminated.

\section*{Acknowledgements}
This work was supported by the Russian Science Foundation grant no.~18-12-00050. We express
gratitude to Dr. G.F.~Gronchi for reviewing this manuscript and providing useful comments.

\appendix

\section{Determining the scan range on a hyperbolic orbit}
\label{sec_hrange}
Let us introduce vector $\bmath W$, which is directed to the ascending node of $\mathcal
O'$ assuming reference $\mathcal O$, and has length $W=\sin I$, where $I$ is mutual
inclination between the orbits. The components of $\mathbf W$ are given in
\citep{BalMik19}. Then the true anomaly of this node in $\mathcal O$ is $\theta_\Omega$,
where
\begin{equation}
 \cos\theta_\Omega = (PW)/W, \quad \sin\theta_\Omega = (QW)/W,
\end{equation}
while $\bmath P$ and $\bmath Q$ are orthogonal unit vectors in the orbital plane
\citep{Baluev05,BalMik19}. The location on the other orbit $\theta_\Omega'$ can be
determined in a similar way through $\bmath P'$ and $\bmath Q'$.

After that let us compute
\begin{align}
r_\pm &= \frac{p}{1\pm e\cos\theta_\Omega}, &r_\pm' &= \frac{p'}{1\pm e'\cos\theta_\Omega'}, \nonumber\\
d_1 &= r_+ - r_+', & d_2 &= r_- - r_-', \nonumber\\
d_3 &= r_+ + r_-', & d_4 &= r_- + r_+', \nonumber\\
d_\Omega &= \min_{r_\pm,r_\pm'>0}(|d_1|,|d_2|,|d_3|,|d_4|),
\label{ind}
\end{align}
where the minimum is considered only among $d_k$ that do not involve negative radii. Notice
that $\theta_\Omega$ or $\theta_\Omega'$ may lie on the imaginary hyperbola branch with
negative $r_+$ or $r_+'$, but in such a case $r_-$ or $r_-'$ would be necessarily positive,
and vice versa. It is also possible that both $r_\pm$ (or $r_\pm'$) appear positive
simultaneously. In \citep{BalMik19} we used only the internodal distances $d_1$ and $d_2$
to compute $d_\Omega$, but $d_3$ and $d_4$ may appear smaller in some cases
\citep{MikBal19}. In the hyperbolic case it is important to take into account $d_{3,4}$,
because it may appear that $d_{1,2}$ both involve imaginary branches.

If $\mathcal O$ is an ellipse, the rest remains the same as in \citep{BalMik19}. Let us
define the quantity $k\geq 0$ and the angle $\varphi$ from
\begin{align}
A^2 = 1 - e^2 \cos^2\theta_\Omega, \quad k = \frac{d_\Omega}{a W A}, \nonumber\\
\sin\varphi = \frac{\sin\theta_\Omega}{A}, \quad \cos\varphi = \sqrt{1-e^2}\, \frac{\cos\theta_\Omega}{A},
\end{align}
then the necessary $u$ range is derived from the inequality
\begin{equation}
e\sin\varphi - k \leq \sin(\varphi-u) \leq e \sin\varphi + k.
\end{equation}
It has three types of solutions discussed in \citep{BalMik19}. The only possible difference
here is due to a refined $d_\Omega$ in~(\ref{ind}).

Whenever $\mathcal O$ is a hyperbola, the matters get more complicated. Then we have
\begin{align}
em - \frac{d_\Omega}{|a|W} \leq m \cosh u + n \sinh u \leq em + \frac{d_\Omega}{|a|W}, \nonumber\\
m=\sin\theta_\Omega, \quad n=\cos\theta_\Omega\sqrt{e^2-1},
\end{align}
and the solution depends on the relationship between $|m|$ and $|n|$.

If $|\tan\theta_\Omega|>\sqrt{e^2-1}$, we put
\begin{align}
A^2 = 1 - e^2 \cos^2\theta_\Omega, \quad k = \frac{d_\Omega}{|a| W A}, \nonumber\\
\tanh\varphi = \sqrt{e^2-1}\, \cot\theta_\Omega,
\end{align}
Then the inequality becomes
\begin{equation}
e\cosh\varphi - k \leq \cosh(\varphi+u) \leq e \cosh\varphi + k.
\end{equation}
This inequality determines two types of ranges for $u$:
\begin{enumerate}
\item If $e\cosh\varphi > 1+k$ then there are two segments surrounding the nodes,
$[\arcosh(e\cosh\varphi-k)-\varphi, \arcosh(e\cosh\varphi+k)-\varphi]$ and
$[-\arcosh(e\cosh\varphi+k)-\varphi, -\arcosh(e\cosh\varphi-k)-\varphi]$.

\item If $e\cosh\varphi \leq 1+k$ then there is a single large segment
$[-\arcosh(e\cosh\varphi+k)-\varphi, \arcosh(e\cosh\varphi+k)-\varphi]$.
\end{enumerate}

Otherwise, if $|\tan\theta_\Omega|<\sqrt{e^2-1}$, we define
\begin{align}
A^2 = e^2 \cos^2\theta_\Omega-1, \quad k = \frac{d_\Omega}{|a| W A}, \nonumber\\
\coth\varphi = \sqrt{e^2-1}\, \cot\theta_\Omega,
\end{align}
and the inequality is
\begin{equation}
e\sinh\varphi - k \leq \sinh(\varphi+u) \leq e \sinh\varphi + k.
\end{equation}
This always defines a single segment $u\in [\arsinh(e\sinh\varphi-k)-\varphi,
\arsinh(e\sinh\varphi+k)-\varphi]$.

Finally, let us consider the degenerate case $|\tan\theta_\Omega|=\sqrt{e^2-1}$, inferring
that one of the nodes coincides with an asymptote of $\mathcal O$. This is a degenerate
case, so we presently do not include it in our code, but still it might be interesting for
the sake of completeness. In this case one of $r_\pm$ is infinite, but the other one
$r_\mp=p/2$ is then finite and positive, so $d_\Omega$ should remain finite. It remains
finite even if this degeneracy is present on the both orbits simultaneously (and hence,
orbits have a common asymptote). We have now $A=0$ and $\varphi=\infty$, so we put $k = (e
d_\Omega)/(|a| W \sqrt{e^2-1})$, and then
\begin{equation}
e - k \leq \exp(\pm u) \leq e + k,
\end{equation}
where the sign coincides with the sign of $\tan\theta_\Omega$. There are two types of
the $u$-range:
\begin{enumerate}
\item If $k>e$ then $\pm u \in [\log(e-k),\log(e+k)]$.
\item If $k\leq e$ then $\pm u \leq\log(e+k)$, and this remains to be the only unbounded
case.
\end{enumerate}

\bibliographystyle{model2-names}
\bibliography{distalghyp}

\begin{thebibliography}{12}
\expandafter\ifx\csname natexlab\endcsname\relax\def\natexlab#1{#1}\fi
\expandafter\ifx\csname url\endcsname\relax
  \def\url#1{\texttt{#1}}\fi
\expandafter\ifx\csname urlprefix\endcsname\relax\def\urlprefix{URL }\fi
\providecommand{\eprint}[2][]{\url{#2}}
\providecommand{\bibinfo}[2]{#2}
\ifx\xfnm\relax \def\xfnm[#1]{\unskip,\space#1}\fi
\bibitem[{Armellin et~al.(2010)Armellin, di~Lizia, Berz and
  Makino}]{Armellin10}
\bibinfo{author}{Armellin, R.}, \bibinfo{author}{di~Lizia, P.},
  \bibinfo{author}{Berz, M.}, \bibinfo{author}{Makino, K.},
  \bibinfo{year}{2010}.
\newblock \bibinfo{title}{Computing the critical points of the distance
  function between two {K}eplerian orbits via rigorous global optimization}.
\newblock \bibinfo{journal}{Celest. Mech. Dyn. Astron.} \bibinfo{volume}{107},
  \bibinfo{pages}{377--395}.
\bibitem[{Baluev and Mikryukov(2019)}]{BalMik19}
\bibinfo{author}{Baluev, R.V.}, \bibinfo{author}{Mikryukov, D.V.},
  \bibinfo{year}{2019}.
\newblock \bibinfo{title}{Fast error-controlling {MOID} computation for
  confocal elliptic orbits}.
\newblock \bibinfo{journal}{\ac} \bibinfo{volume}{27}, \bibinfo{pages}{11--22}.
\bibitem[{Baluyev and Kholshevnikov(2005)}]{Baluev05}
\bibinfo{author}{Baluyev, R.V.}, \bibinfo{author}{Kholshevnikov, K.V.},
  \bibinfo{year}{2005}.
\newblock \bibinfo{title}{Distance between two arbitrary unperturbed orbits}.
\newblock \bibinfo{journal}{Celest. Mech. Dyn. Astron.} \bibinfo{volume}{91},
  \bibinfo{pages}{287--300}.
\bibitem[{Dybczy\'{n}ski et~al.(1986)Dybczy\'{n}ski, Jopek and
  Serafin}]{Dybczynski86}
\bibinfo{author}{Dybczy\'{n}ski, P.A.}, \bibinfo{author}{Jopek, T.J.},
  \bibinfo{author}{Serafin, R.A.}, \bibinfo{year}{1986}.
\newblock \bibinfo{title}{On the minimum distance between two {K}eplerian
  orbits with a common focus}.
\newblock \bibinfo{journal}{Celest. Mech.} \bibinfo{volume}{38},
  \bibinfo{pages}{345--356}.
\bibitem[{Gronchi(2002)}]{Gronchi02}
\bibinfo{author}{Gronchi, G.F.}, \bibinfo{year}{2002}.
\newblock \bibinfo{title}{On the stationary points of the squared distance
  between two ellipses with a common focus}.
\newblock \bibinfo{journal}{SIAM J. Sci. Comput.} \bibinfo{volume}{24},
  \bibinfo{pages}{61--80}.
\bibitem[{Gronchi(2005)}]{Gronchi05}
\bibinfo{author}{Gronchi, G.F.}, \bibinfo{year}{2005}.
\newblock \bibinfo{title}{An algebraic method to compute the critical points of
  the distance function between two {K}eplerian orbits}.
\newblock \bibinfo{journal}{Celest. Mech. Dyn. Astron.} \bibinfo{volume}{93},
  \bibinfo{pages}{295--329}.
\bibitem[{Gronchi and Niederman(2020)}]{GronchiNiederman20}
\bibinfo{author}{Gronchi, G.F.}, \bibinfo{author}{Niederman, L.},
  \bibinfo{year}{2020}.
\newblock \bibinfo{title}{On the nodal distance between two {K}eplerian
  trajectories with a common focus}.
\newblock \bibinfo{journal}{Celest. Mech. Dyn. Astron.} \bibinfo{volume}{132},
  \bibinfo{pages}{5}.
\bibitem[{Guzik et~al.(2019)Guzik, Drahus, Rusek, Waniak, Cannizzaro and
  Pastor-Marazuela}]{Guzik19}
\bibinfo{author}{Guzik, P.}, \bibinfo{author}{Drahus, M.},
  \bibinfo{author}{Rusek, K.}, \bibinfo{author}{Waniak, W.},
  \bibinfo{author}{Cannizzaro, G.}, \bibinfo{author}{Pastor-Marazuela, I.},
  \bibinfo{year}{2019}.
\newblock \bibinfo{title}{Initial characterization of interstellar comet
  {2I/Borisov}}.
\newblock \bibinfo{journal}{Nat. Astron.} \bibinfo{volume}{4},
  \bibinfo{pages}{53--57}.
\bibitem[{Hedo et~al.(2018)Hedo, Ru\'{i}z and Pel\'{a}ez}]{Hedo18}
\bibinfo{author}{Hedo, J.M.}, \bibinfo{author}{Ru\'{i}z, M.},
  \bibinfo{author}{Pel\'{a}ez, J.}, \bibinfo{year}{2018}.
\newblock \bibinfo{title}{On the minimum orbital intersection distance
  computation: a new effective method}.
\newblock \bibinfo{journal}{MNRAS} \bibinfo{volume}{479},
  \bibinfo{pages}{3288--3299}.
\bibitem[{Kholshevnikov and Vassiliev(1999)}]{KholshVas99}
\bibinfo{author}{Kholshevnikov, K.}, \bibinfo{author}{Vassiliev, N.},
  \bibinfo{year}{1999}.
\newblock \bibinfo{title}{On the distance function between two {K}eplerian
  elliptic orbits}.
\newblock \bibinfo{journal}{Celest. Mech. Dyn. Astron.} \bibinfo{volume}{75},
  \bibinfo{pages}{75--83}.
\bibitem[{Mikryukov and Baluev(2019)}]{MikBal19}
\bibinfo{author}{Mikryukov, D.V.}, \bibinfo{author}{Baluev, R.V.},
  \bibinfo{year}{2019}.
\newblock \bibinfo{title}{A lower bound of the distance between two elliptic
  orbits}.
\newblock \bibinfo{journal}{Celest. Mech. Dyn. Astron.} \bibinfo{volume}{131},
  \bibinfo{pages}{28}.
\bibitem[{Sitarski(1968)}]{Sitarski68}
\bibinfo{author}{Sitarski, G.}, \bibinfo{year}{1968}.
\newblock \bibinfo{title}{Approaches of the parabolic comets to the outer
  planets}.
\newblock \bibinfo{journal}{Acta Astronomica} \bibinfo{volume}{18},
  \bibinfo{pages}{171--195}.

\end{thebibliography}







\end{document}